\documentclass{rQUF2e}
\usepackage{mathtools}
\usepackage[hang,small,bf]{caption}
\usepackage[subrefformat=parens]{subcaption}
\newcommand{\MARU}[1]{{\ooalign{\hfil#1\/\hfil\crcr\raise.167ex\hbox{\mathhexbox20D}}}}

\begin{document}
\title{{\textit{Improving Regression-Based Event Study Analysis Using a Topological Machine-Learning Method}}}% Force line breaks with \\
\author{Takashi Yamashita$^{\ast}$$\dag$\thanks{$^\ast$Corresponding author.
Email: tak20mt06@gmail.com} and 
Ryozo Miura${\ddag}$\thanks{$^\ast$Coauthor. Email: ryozomiura@gmail.com}\\
\affil{$\dag$Tokyo University of Science, 1-3, Kagurazaka, Shinjyuku-ku,
Tokyo 162-0825, Japan\\
$\ddag$Tohoku University, 27-1 Kawauchi, Aoba-ku,
Sendai-shi, Miyagi 981-8576, Japan}} 
\date{\today}
\maketitle
\section{Abstract} \label{Sec:Abstract}
This paper introduces a new correction scheme to a conventional regression-based event study method: a topological machine-learning approach with a self-organizing map (SOM). We use this new scheme to analyze a major market event in Japan and find that the factors of abnormal stock returns can be easily identified and the event-cluster can be depicted. We also find that a conventional event study method involves an empirical analysis mechanism that tends to derive bias due to its mechanism, typically in an event–clustered market situation. \par
We explain our new correction scheme and apply it to an event in the Japanese market --- the holding disclosure of the Government Pension Investment Fund (GPIF) on July 31, 2015.
\begin{keywords}
event clustering; 
self-organizing map; 
abnormal return misdetection; 
linear regression market model 
\end{keywords}
\section{Introduction} \label{Sec:Introduction}
Event study analysis measures the effects of an economic event on the value of firms. This type of analysis is one of the most important methodologies in finance research. 
Empirical research based on event studies has been applied to areas of corporate finance, such as M\&As and funding, and has greatly influenced business practices including insider trading certification and legal regulations. \par
The impact of an event must be verified using objective and reproducible methods. 
Event studies use statistical hypothesis tests for verification purposes. An abnormal return is the most important indicator used in hypothesis testing.
An abnormal return is the difference between the actual return and the expected return of a security obtained with some market model that corresponds to the estimation error of the model.
A linear regression factor model is the most basic tool and is commonly used in empirical analysis. The return on any security can be expressed as a linear combination of factor returns in this type of model.  
The capital asset pricing model (CAPM) is a typical one-factor model with a formula that calculates the expected return on a security based on its level of risk, ease of understanding, and price fluctuation mechanism.   
One of the major reasons a linear regression market model is used in many empirical studies is its ability to identify with a simple linear regression algorithm. \par
From a statistical perspective, an event study detects the bias of the estimation error in the linear algebra model.
This suggests two drawbacks to this type of event study.
This method does not explain whether the observed abnormal returns are the result of an event of interest. This drawback occurs because the influence of the event is indirectly expressed as a relatively large error of the estimated price in the event period.
The probability of incorrect detection risk increases in the event of major fluctuations in the pricing mechanism of return for the underlying security. This is because the linear regression model assumes linear relations and stationarity between the explanatory and explained variables.
To compensate for these problems, we aggregate abnormal returns, and the comprehensively deduce the statistics in an event study. 
However, it is not always possible to aggregate abnormal returns.
The analysis that aggregates abnormal returns assumes that cross-sectional correlations between all securities are zero.
If the event windows overlap, this assumption cannot be applied.
Event clustering is a major problem associated with event studies that researchers have discussed\cite{bernard1987cross,li2002semiparametric,brewer2003value,brewer2003does}. 
This problem can be treated using one of two methods.
The first method is analyzing the return of securities without aggregating. Abnormal returns are estimated using a multivariate regression model with dummy variables for the event day.   
This approach has some advantages using an alternative hypothesis where positive or negative abnormal returns for securities can be accommodated. On the other hand, this approach often provides minimal statistical power against economically reasonable alternatives. The other problem is not applicable where event dates are completely overlapping.  
The second method aggregates a portfolio by event time and applies a single security analysis to the portfolio. This approach is easy to analyze; however the result strongly depends on the security selection.   
In either case, the methods have the same structure by which the influence of an event is indirectly detected as an extension of estimation error for return using a market model.
In principle, both methods cannot identify events that cause abnormal returns.
Additional analysis is necessary to separate events causing the abnormal returns in event study situations.
%%%%
For this purpose, we focus on a topographic data analysis (TDA), which is an approach to the analysis of datasets using techniques from topology. TDA estimates the "shape" that essentially characterizes the analysis object from the data, and enables identification and classification based on the shape without intervention in the analysts' judgment. 
TDA is a relatively new field of research with few examples of application in the finance field.
A generative topographic map (GTM) with an improved self-organizing map (SOM) architecture is a method that estimates the phase structure of data. Since GTM can hold distance information between data, data similarity can be expressed by distance. 
SOM distance information does not necessarily reflect similarity, but it has a property whereby expanding the region where information is dense facilitates the identification of securities, which is useful for event studies. 
GTM is calculated using likelihood maximization, which assumes a specific probability distribution for the structure of the analysis object. As we describe later, we use various input data subjected to nonlinear transformation in this analysis. Therefore, the assumption of a specific probability distribution may result in an incorrect result. For these reasons, we used SOM instead of GTM.
%%%%
We focus on SOM, which is a machine learning process, and propose a new methodology.
SOM is a type of artificial neural network that was introduced by Kohonen \cite{kohonen1998self} in the 1970s for computational abstraction building on biological models of neural network systems. 
SOM projects high dimensional input data to low dimensional space holding their topological relations using an unsupervised machine learning method.
The term ``topology'' refers to a specific mathematical idea whereby elements of a set relate spatially to each other. 
SOM cannot hold distance information between data elements, but it does hold phase information. Therefore, this property is useful for event identification. Traditional calculation methods, such as principal component analysis, identify events while reducing the dimensions of input data but erase information that is not relatively important.
This feature of SOM is also useful for understanding observation as one system.
The input data vectors are composed of abnormal returns in the event window and quantitative information related to other events. 
The quantitative information includes data on the characteristics and reliability of the market model of each security, which is useful for estimating the misdetection of abnormal returns and understanding market mechanisms.
Therefore, it is possible to separate the influence of various events. \par
We present an empirical analysis using this methodology for cases where many events occur at once in the \ref{Sec:Empirical Analysis and discussion}. The results of this analysis are compared with those of traditional event studies and show the validity of the new methodology.
\section{Methodology} \label{Sec:Methodology}
In this paper, we introduce a proposed new event study methodology for event clustering.
We modify the traditional event study scheme in this section.
According to Mackinlay\cite{campbell1997econometrics,mackinlay1997event}, a brief analytical process in the traditional event study is as follows:
\begin{enumerate}
 \item Step1: Define the event.
 \item Step2: Determine the selection criteria of the firm in the study.
 \item Step3: Develop the market model to calculate abnormal returns.
 \item Step4: Design a testing framework for the abnormal returns.
 \item Step5: Present and interpret the empirical results.
\end{enumerate} 
We change Step 2 and Step 4, and these changes affect Step 5.
In traditional schemes, or Step 2, the securities to be tested are selected according to the content of the event. There is no need to screen out using our new methodology.
Some type of statistical testing framework for abnormal returns that defines the null hypothesis is the traditional methodology of Step 4. \par
The null hypothesis $(H_{0})$ thus maintains that there are no cumulative abnormal returns (CARs) within the event window, whereas the alternative hypothesis $(H_{1})$ suggests the presence of CARs within the event window. Formally, the testing framework reads as follows:
\begin{align*}
H_{0}: CAR = 0, \\ 
H_{1}: CAR \neq 0, \\
\intertext{auxiliary hypothesis regarding abnormal returns (ARs) defined as}
H_{0}': AR = 0, \\
H_{1}': AR \neq 0.
\end{align*}
$H_{0}$ is rejected by selecting significance level $\theta$. Common values are $\theta = 5\%$ or 1\%. We selected 5\% as $\theta$ in this study.
For our new methodology, we create a data set to be inputted to the SOM.
This is a multidimensional data set containing values that vary depending on the attributes of each security, AR, CAR, and the event occurrence information.
These data are standardized by standard deviation for each dimension.
The SOM algorithm maps high dimensional observation space to low dimensional latent space.
The latent space is made two-dimensional and divided into a hexagonal lattice in this analysis.
Each lattice (node) has a vector (reference vector) of the same dimension as the observation space.
The SOM learning is simple competitive learning such as "winner takes all". 
The SOM algorithm finds the best matching unit (BMU) by calculating the Euclid distance between the input vector and the weight vector of each node.
Each vector in the BMU's neighborhood is adjusted to become more like the BMU based on the distance from the BMU.
The learning rate is also an exponential decay on each iteration.
Both stochastic on-line and deterministic batch algorithms were designed for SOM. 
This is one of the vector quantization’s for which a lossy data compression method does not degrade the data.
Note that the BMU modifies the input data vector of each security. 
The results obtained by SOM analysis are helpful in assessing the overall tendency of the observations. On the other hand, the results lack statistical precision. We must reconfirm the t-values of individual securities for accurate statistical testing.
There are two algorithms for SOM, one is online learning, and the other is batch learning.   
An on-line algorithm is designed to model some plastic features of the human brain, and the learning result depends on the data input order. This feature is a drawback for data analysis. We selected batch learning algorithm for this analysis.\cite{vesanto2000clustering}  
Each security is labeled with a node with the most similar reference vector.
If an abnormal return occurs due to a specific event, the area strongly related to the event must match the area where the abnormal return occurred.
Using this property, we can infer the relationship between abnormal returns and events.
\section{Empirical Analysis and discussion} \label{Sec:Empirical Analysis and discussion}
In this section, we analyze an event study using our new methodology.
The event is the holding disclosure on July 31, 2015, of the Government Pension Investment Fund (GPIF), which is the world's largest public pension fund in Japan.
On July 29, 2016, GPIF released all stock holdings as of the end of March 2015.\par
\begin{figure}[htbp]
  \centering %
    \includegraphics[scale=0.8]{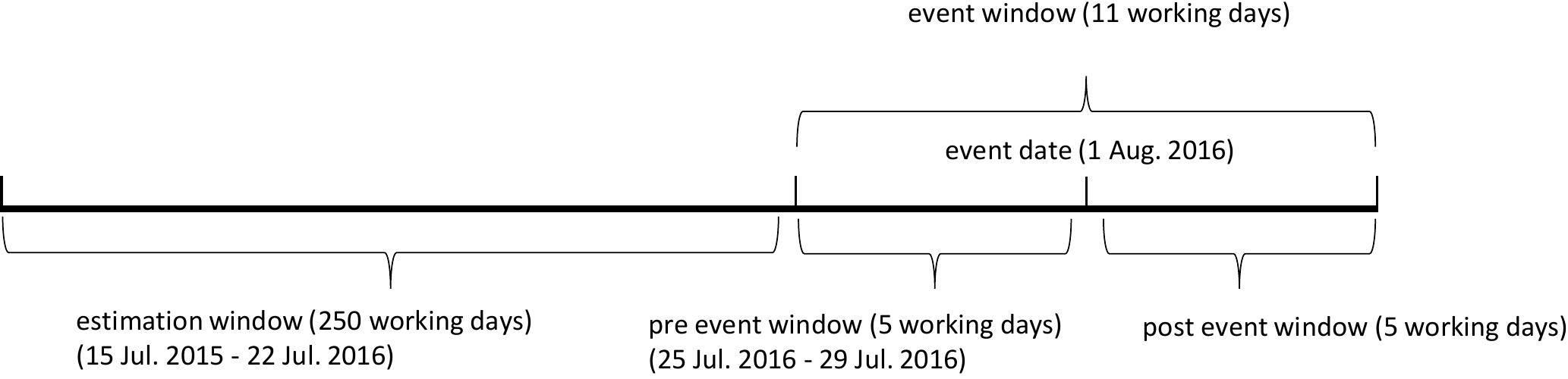}
    \caption{Timeline of event study of this analysis} %
    \label{Fig:TimeLine} %
\end{figure}
\begin{figure}[htbp]
  \centering %
    \includegraphics[scale=0.9]{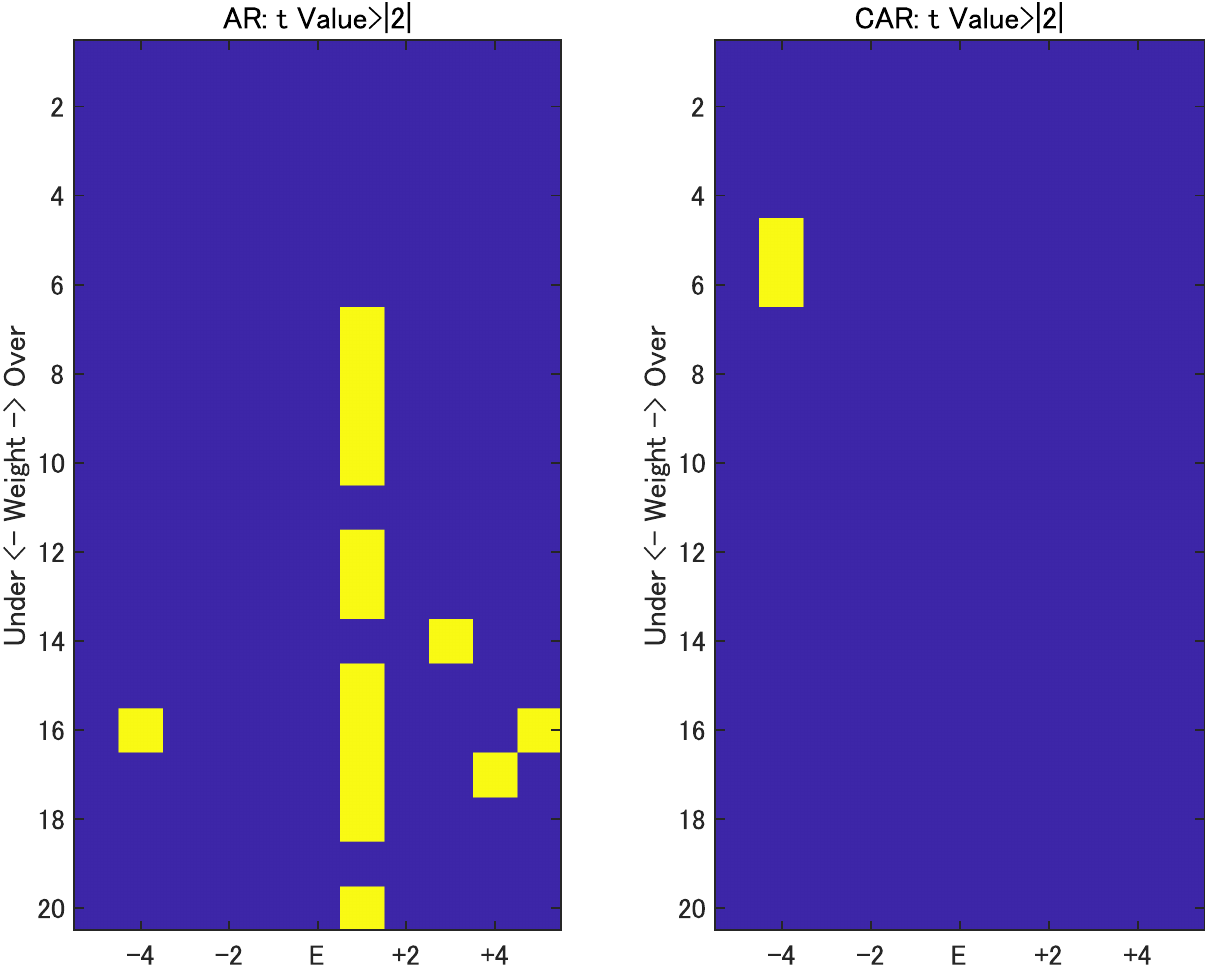}
    \caption{Results of detected AR and CAR of sorted portfolio with GPIF's modified active weight. Yellow represents detected AR (or CAR), and blue represents undetected AR (or CAR). Vertical axis indicates the order of modified active weight, horizontal axis indicates the interval from the event day.} %
    \label{Fig:Quantile20} %
\end{figure}
\begin{figure}[htbp]
  \centering %
    \includegraphics[scale=0.9]{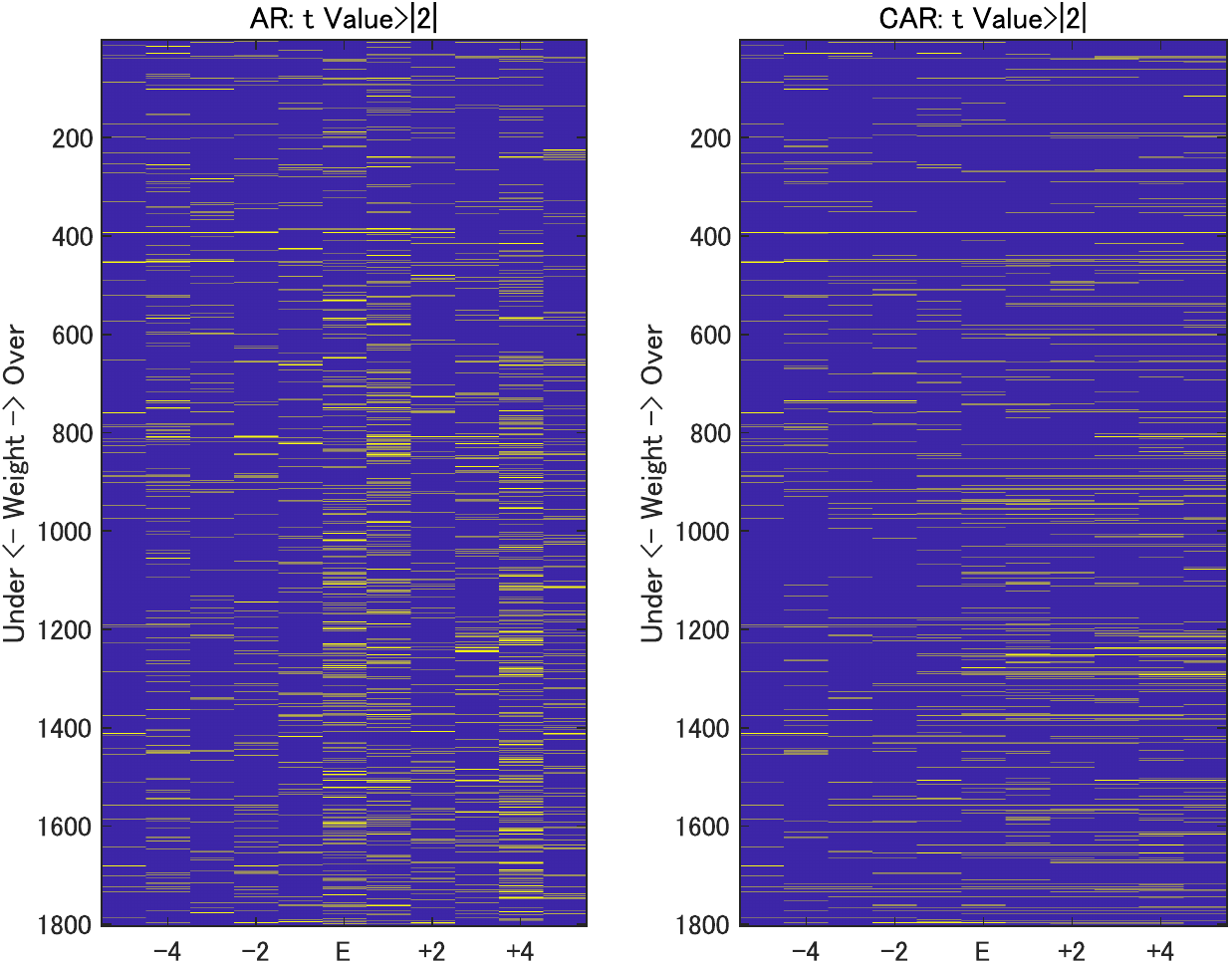}
    \caption{Results of detected AR and CAR of individual securities with GPIF's modified active weight. Yellow represents detected AR (or CAR), and blue represents undetected AR (or CAR). Vertical axis indicates the order of modified active weight, horizontal axis indicates the interval from the event day.} %
    \label{Fig:NoQuantile} %
\end{figure}

% 35178400/580720220
GPIF is a huge fund holding approximately 6\% of the Japanese stock market. 
%%%%%
Most of the world's leading pension funds, including the GPIF, use the performance of each country's stock market as a criterion for equity investment by country. 
This portfolio for calculating performance is called a benchmark index. 
The benchmark index is composed of the stocks listed on the country’s stock market, and the investment weight (allocation ratio) is the market capitalization ratio for the whole listed market. If an investment manager does not have a specific market view on individual securities, the manager has a portfolio with the same weight as the benchmark index. This is because the performance of the investment portfolio is linked to that of the country's stock market. 
Therefore, it is possible to estimate the market view of the investment manager by observing the deviation from the benchmark index of the securities that compose the portfolio. The deviation ratio from the benchmark index is called an ``”activeweight.”
GPIF must verify whether this divergence information would affect the market price formation mechanism. This is because GPIF is prohibited from affecting market prices by law.
%%%%%
Market participants can determine the active weight of equity based on the holding information of GPIF.
%%%%%
%%%%%
Some consider that active weights reflect the evaluation of GPIF for each equity, and this information may affect the investment decisions of market participants. For this reason, GPIF has never published its holding information.
Since this event has only one window, the event study in this case is considered event clustering.
Additionally, Table \ref{Tbl:Events} shows that, at this time, many other events occurred that may have affected the market.
At the end of June 2016, a Federal Open Market Committee (FOMC), a committee within the US Federal Reserve System, was held in the United States, and a Monetary Policy Meeting was held in Japan. The results were announced on June 27 and June 29, respectively.
The election of the heads of Tokyo was held on July 31.
From July 25 to August 7, the event window, the settlement of 1,322 companies was announced. These companies account for approximately two-thirds of the analyzed securities. On July 29, 309 companies' earnings announcements were concentrated. These companies accounted for approximately 16\% of the total companies. 
The analysis was conducted according to the methodology described in the \ref{Sec:Methodology}. 
The event window lasted for 11 business days from July 25 to August 8, 2016. The estimation windows lasted 250 business days from July 22, 2016, to July 24, 2015. 
The timing sequence is indicated on the timeline in Fig. \ref{Fig:TimeLine}.
The securities analyzed represent the first section of the Tokyo Stock Exchange. However, equities for which the market price could not be acquired due to issues such as consolidation and delisting were excluded.
For comparison, we implemented an event study using a traditional method.
For event clustering, we created sorted portfolios of equal weights. The criterion for sorting was the deviation from the market average of the securities holding ratio of GPIF.
The deviation $W_{\mathrm{active}}$ is defined below: 
%AW = W_market – W_gpif
\begin{equation}
W_{\mathrm{active}} = W_{\mathrm{market}} - W_{\mathrm{GPIF}}
\end{equation} \label{Eq:ActiveW}
$W_{\mathrm{active}}$ has large‐sized capital stocks bias. To avoid this bias, we used the standardized weight $W_{\mathrm{mod_active}}$ described below for each security as a criterion.
%mAW = AW/W_market
\begin{equation} \label{Eq:modActiveW}
W_{\mathrm{modactive}} = W_{\mathrm{active}} / W_{\mathrm{market}}
\end{equation} 
The market model for both the traditional and proposed methods is a single factor model for which the explanatory variable of the security return is a benchmark of the market as shown below:
\begin{equation} \label{Eq:MarketModel}
r_{i,t} = \alpha_{i} + \beta_{i} r_{\mathrm{market},t} + \nu_{i,t}
\end{equation} 
The null hypothesis of no mean event effect reduces to
\begin{equation}
H_{0}:\mu = 0~~~ (H'_{0}:\mu = 0),
\end{equation}
where $\mu$ is the expectation of CAR (AR). Using the p-value method, we rejected $H_{0}$ at the $\theta=0.05$ level. The corresponding t-value is approximately $2.1$.
Fig. \ref{Fig:Quantile20} shows the results of detected AR and CAR for a sorted portfolio composed of equal weight securities with GPIF's modified active weight according to \eqref{Eq:modActiveW}.
The yellow portfolios do not reject the null hypothesis $H_{0}$. 
The figure shows that we cannot conclude that there is no effect of holding disclosure because $H_{0}$ cannot be rejected. However, this result is not natural because there is no obvious relationship between the amount of divergence of active weights and the frequency of abnormal returns.
Fig. \ref{Fig:NoQuantile} shows the results of detected AR and CAR of sorted individual securities with GPIF's modified active weight. Table \ref{Tbl:Percent} indicates the percentage of securities that rejected hypothesis $H_{0}$ for all (1804) securities. 
Table \ref{Tbl:Percent} shows that securities with abnormal returns increased after the event day. This result indicates that the market price mechanism changed significantly around the event day.
This finding and the results in Fig.\ref{Fig:NoQuantile} imply that abnormal return detection does not depend on the modified active weight.
However, we cannot conclude that the GPIF holdings disclosure event did not affect the market pricing mechanism.
This is because there is a possibility that the abnormal return occurred due to other events, such as a financial statements announcement, which were mixed with the effect of the holdings disclosure making the effects difficult to separate. 
Using only a traditional analysis, it is not possible to evaluate the influence of a specific event. \par
We prepared data sets to separate the effect of events using SOM analysis.
These data sets are used to correct two drawbacks of traditional event study using the linear regression type market model noted in the \ref{Sec:Introduction}.
The first added data set is used to isolate the effect of each event that occurred as a duplicate during the event window.
From Table \ref{Tbl:Events}, a schedule diagram of the event window, we paid attention to two events that affect the price formation mechanism of the market during the event window. One event is the Bank of Japan's Monetary Policy Meeting on July 28 and 29, and the other is the financial statements announcements for each security around the event day.  
To evaluate the first event, we used the correlation coefficient $\rho$ between the annual interest rate and the stock price return as a variable to explain the events of the BOJ's Monetary Policy meeting. This is because the central bank manipulates short-term interest rates, which strongly influence the bank's revenue.
To evaluate the second event, we selected the number of days from the event day to the financial statements announcements as a variable to explain their impact.
The period between the event day and the financial statements announcement for each security was converted using the following formula:
%tanh(1./Z.KessDist')
\begin{equation}
% \tanh{(\frac{1}{D_{\mathrm{ED}}-D_{\mathrm_{AFS}}})}
\Omega = \tanh{(\frac{1}{D_\mathrm{event}-D_\mathrm{annaunce}})}
\end{equation}.
The additional data vectors that we use to evaluate the quality of the market model for each security are given by DWR, $\hat{\alpha}$ and $\hat{\beta}$ where DWR is Durbin-Watson ration of residuals, and $\hat{\alpha}$ and $\hat{\beta}$  are estimated intercept and regression coefficients.
The data set vector $Q$ for the analysis is given by:
\begin{align}
Q = 
\begin{bmatrix}
mAW_{1} & mAW_{2} & \cdots & mAW_{N} \\
tCAR_{-5d,1} & tCAR_{-5d,2} & \cdots & tCAR_{-5d,N} \\
tCAR_{-4d,1} & tCAR_{-4d,2} & \cdots & tCAR_{-4d,N} \\
\vdots & \vdots & \vdots & \vdots \\
tCAR_\mathrm{ed,1} & tCAR_\mathrm{ed,2} & \cdots & tCAR_\mathrm{ed,N} \\
\vdots & \vdots & \vdots & \vdots \\
tAR_\mathrm{+5d,1} & tAR_\mathrm{+5d,2} & \cdots & tAR_\mathrm{+5d,N} \\
tAR_\mathrm{-5d,1} & tAR_\mathrm{-5d,2} & \cdots & tAR_\mathrm{-5d,N} \\
tAR_\mathrm{-4d,1} & tAR_\mathrm{-4d,2} & \cdots & tAR_\mathrm{-4d,N} \\
\vdots & \vdots & \vdots & \vdots \\
tAR_\mathrm{0,1} & tAR_\mathrm{0,2} & \cdots & tAR_\mathrm{0,N} \\
\vdots & \vdots & \vdots & \vdots \\
tAR_\mathrm{+5d,1} & tAR_\mathrm{+5d,2} & \cdots & tAR_\mathrm{+5d,N} \\
\rho_\mathrm{1y,1} & \rho_\mathrm{1y,2} & \cdots & \rho_\mathrm{1y,N} \\
\Omega_{1} & \Omega_{2} & \cdots & \Omega_{N} \\
DWR_{1} & DWR_{2} & \cdots & DWR_{N} \\
\hat{\alpha}_{1} & \hat{\alpha}_{2} & \cdots & \hat{\alpha}_{N} \\
\hat{\beta}_{1} & \hat{\beta}_{2} & \cdots & \hat{\beta}_{N}
\end{bmatrix},
\end{align}
Here, $mAW$ represents the modified active weight, subscript represents the number of the sample, and 1,804 securities are applied for this analysis($N=1804$). The first subscript for each CAR and AR represents the business interval day from the event day, and the second subscript represents the number of securities. \par
\begin{figure}[htbp]
 \begin{minipage}[b]{0.9\linewidth}
  \centering
  \includegraphics[keepaspectratio, scale=0.6]{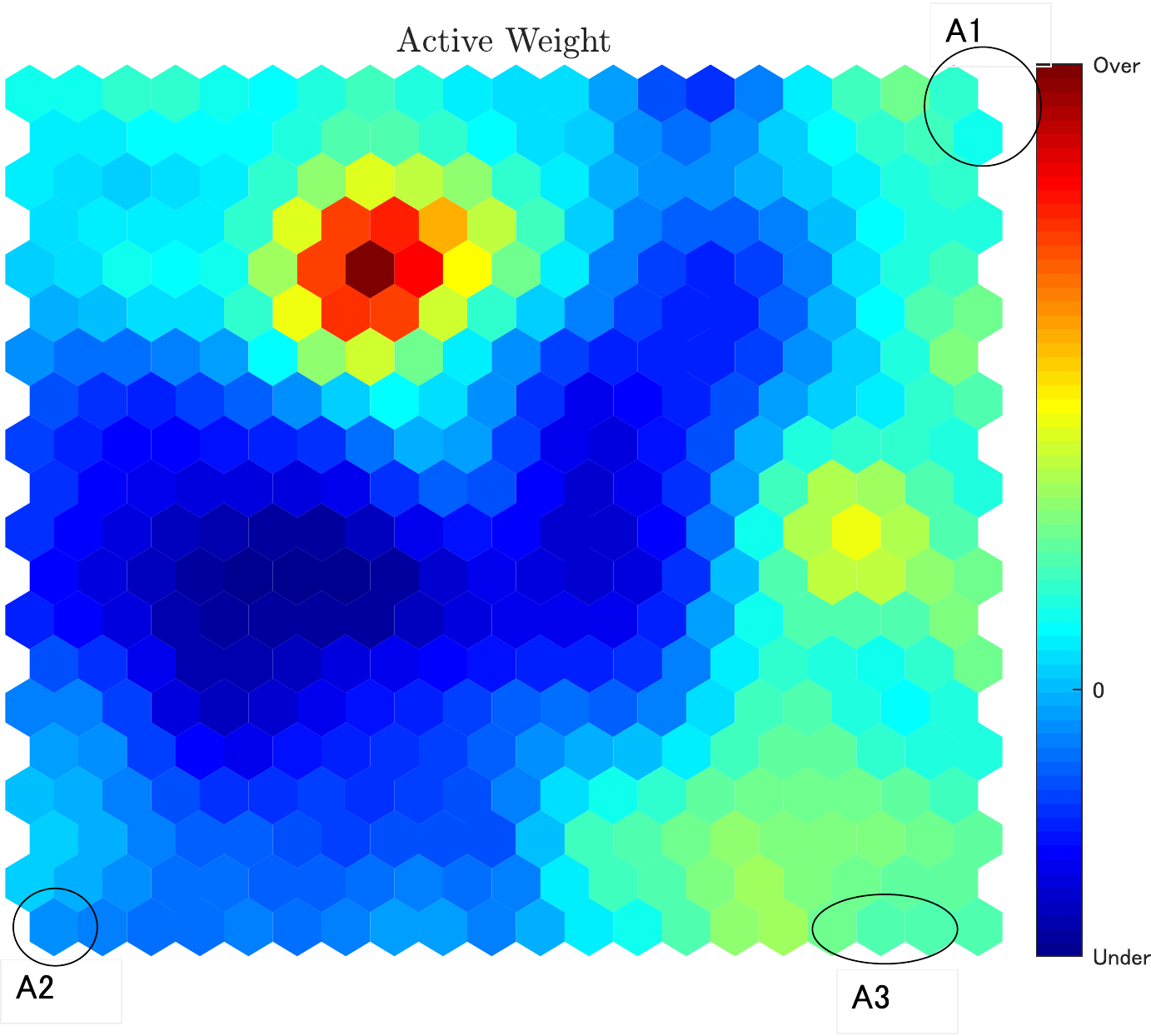}
  \subcaption{SOMs for GPIF Active Weight.}\label{Fig:SOM_ActiveWeight}
 \end{minipage}\\
\begin{minipage}[b]{0.9\linewidth}
  \centering
  \includegraphics[keepaspectratio, scale=0.6]{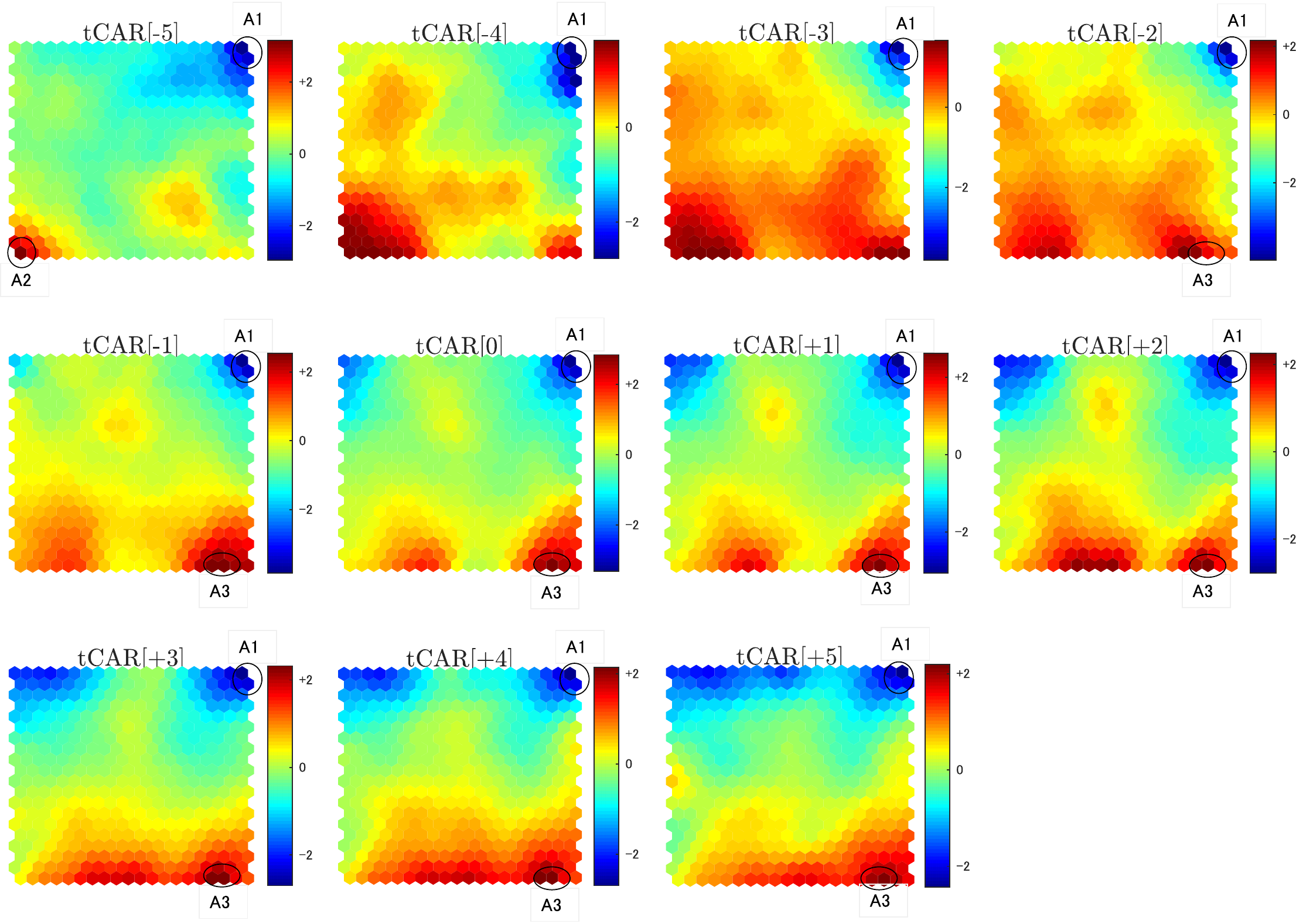}
  \subcaption{SOMs for GPIF Active Weight vs. Cumulative Abnormal Return in Event Window.}\label{Fig:SOM_CAR}
 \end{minipage}\\
 \caption{Self-organization MAPs (SOMs) for GPIF's Active Weight and CAR .}\label{Fig:SOM_ActiveWeight_CAR}
\end{figure}

\begin{figure}[htbp]
 \begin{minipage}[b]{0.9\linewidth}
  \centering
  \includegraphics[keepaspectratio, scale=0.6]{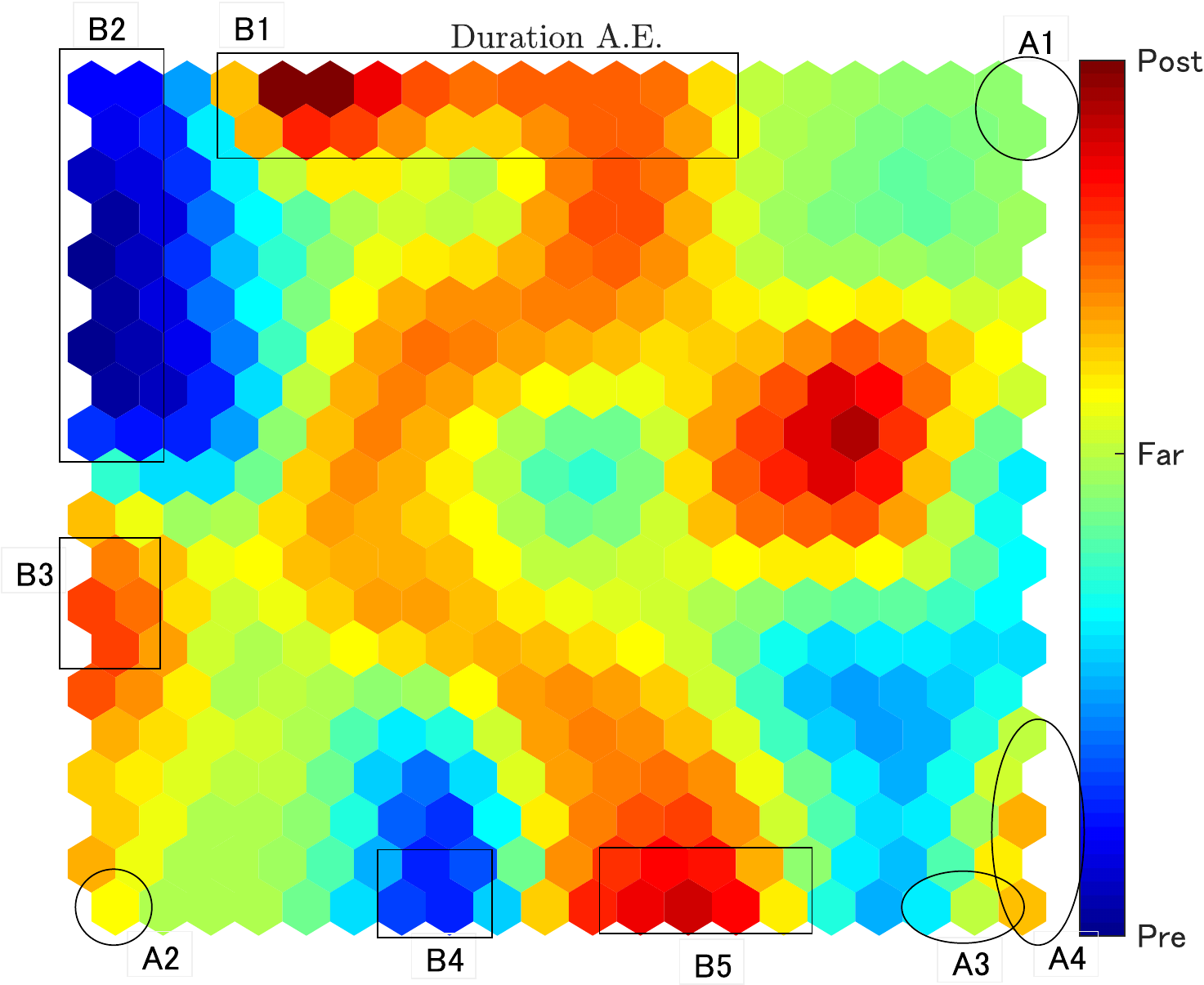}
  \subcaption{SOMs for announcement date of firm's earnings around the event day}\label{Fig:SOM_DurationAE}
 \end{minipage}\\
 \begin{minipage}[b]{0.9\linewidth}
  \centering
  \includegraphics[keepaspectratio, scale=0.6]{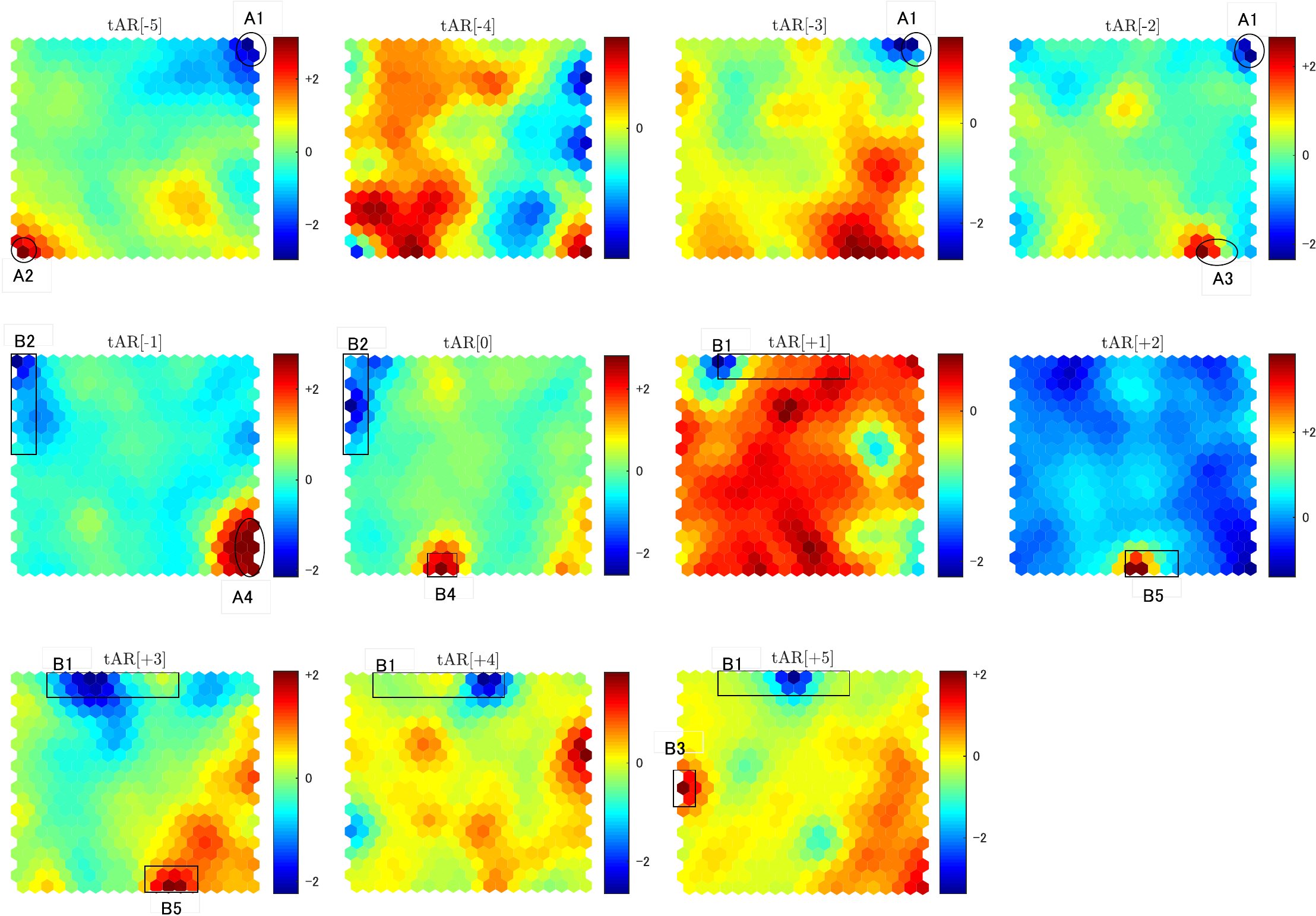}
  \subcaption{SOMs for abnormal return in event window}\label{Fig:SOM_AR}
 \end{minipage}\\
 \caption{Self-organization MAPs (SOMs) for announcement date of a firm's earnings around the event day and AR.}\label{Fig:SOM_ActiveWeight_AR}
\end{figure}

\begin{figure}[htbp]
  \begin{minipage}[b]{1.0\linewidth}
  \centering
  \includegraphics[keepaspectratio, scale=0.6]{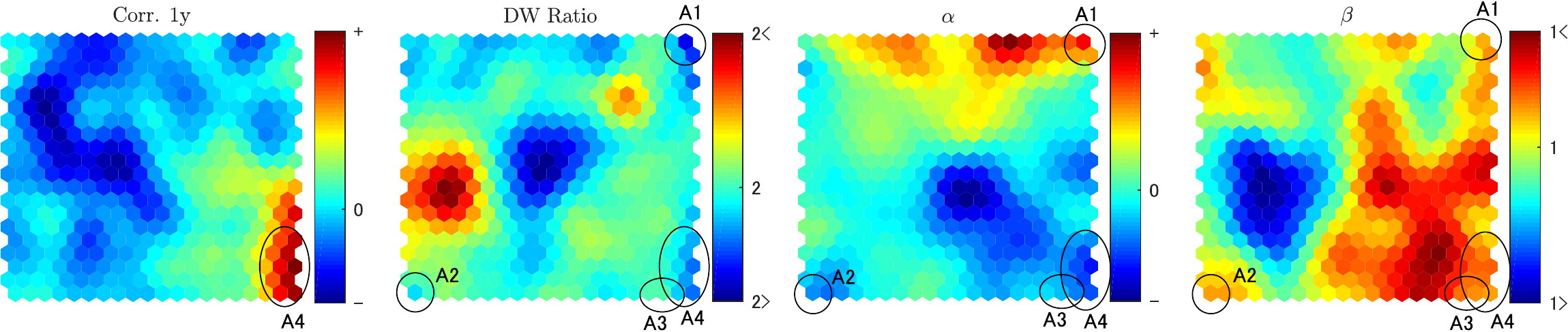}
  \subcaption{SOMs for (1)Pearson’s product moment correlation coefficients between equity return and interest rate, (2)Durbin-Watson ratio for residuals of each firm's market model, (3)intercepts, and (4)regression coefficients for each firm's market model in estimate window.}
  \label{Fig:SOM3}
 \end{minipage}
 \caption{Self-organization maps (SOMs) for the analysis.}\label{Fig:SOM}
\end{figure}

  
The SOMs were obtained using the batch learning algorism (see Appendix \ref{apdx:A}). We take $I=J=20$ ($20 \times 20$ cells map). The learning parameters are shown below:
\begin{equation*}
\lambda_{\mathrm{init}} = 0.9, \xi_{\mathrm{init}}=0.001, T = 2000
\end{equation*}
%
%The learning result for $Q$ is shown in Fig.\ref{Fig:SOM1}, Fig.\ref{Fig:SOM2} and Fig.\ref{Fig:SOM3}.
The learning result for $Q$ is shown in Fig.\ref{Fig:SOM} 。
The SOM algorithm maps all the securities in either lattice on all maps. The relative positional relationship of each security is the same on all maps.
These maps are colored according to the hierarchical value of each variable; that is, they are heat maps.
Fig. \ref{Fig:SOM_ActiveWeight} indicates the map colored according to the quantities of the GPIF holdings active weight.
Maps from tCAR[-5]  to tCAR[+5] in Fig. \ref{Fig:SOM_CAR} and from tAR[-5]  to tAR[+5] in Fig. \ref{Fig:SOM_AR} indicate the cumulative abnormal returns and abnormal returns for all securities, respectively.
Securities that failed to reject the null hypothesis H0 at a significance level of 5\% are included in the colored cells and correspond to an absolute t-value of 2 or more.
We discern three lattices in regions A1, A2, and A3 on the maps in Fig. \ref{Fig:SOM_CAR} from tCAR[-5] to tCAR[+5] include securities that cannot reject the null hypothesis $H_{0}$.
%%%%%%%%%%%%%%%%%%%%%%%%%%%%%%%%%%%
The submap ``Active Weight'' shows that we cannot observe any dependence between the active weight of the GPIF holdings and cumulative abnormal returns. 
%%%%%%%%%%%%%%%%%%%%%%%%%%%%%%%%%
However, according to this result, it would be wrong to conclude that the holdings disclosure of GPIF has no connection with the market pricing mechanism.   
We must identify the cause of abnormal returns in regions A1, A2, and A3. \par
Earnings announcements have an influence on securities’ prices. 
Fig.\ref{Fig:SOM_DurationAE} indicates the relationship between earning announcements and abnormal returns, and the submaps are colored according to the duration between the event day and the earnings announcement.   
"tAR[-5]" to "tAR[+5]" in Fig.\ref{Fig:SOM_AR} show the distribution of securities with an abnormal return on each day. We can identify regions from B1 to B5 with rectangles colored red and blue on submap "Duration E.A." 
We find that most abnormal returns are explained by earnings announcements.\par
On the other hand, for elliptical regions where detecting cumulative abnormal returns in regions A1, A2, and A3 are disaccorded with the rectangular regions from B1 to B5, it is not appropriate to consider that earnings announcements are the cause of cumulative abnormal returns.
A part of the elliptical region A4 overlaps with region A3 and has red cells and green cells. The relationship between abnormal returns and earnings announcements is not clear since green cells show that the earnings announcement day is not close to the event window. Fig. \ref{Fig:SOM3} is a set of submaps useful for identifying cumulative abnormal returns on regions A1 to A3.
The submap "Corr. 1y" indicates the distribution of securities considering the strength of correlation between daily changes in short-term interest rates in Japan and equity returns. The submap "DW Ratio" indicates the Durbin-Watson ratio for the residuals of each securities market model. Submap "$\alpha$" and "$\beta$" indicate their intercepts and regression coefficients.
The elliptical region A4 of submap "tAR[-1]"  in Fig. \ref{Fig:SOM_AR} agrees with the highly correlated area of submap "Corr.1y" in Fig. \ref{Fig:SOM3}.
Five cells in region A4 have 47 securities, 44 of which are banks.

Due to the Bank of Japan’s Monetary Policy Meeting held on July 28 and 29 and an unchanged policy interest rate as of July 29 that went against many market participants predictions of additional monetary easing, banks’ stock prices jumped.
Therefore, we consider the cause of abnormal returns in region A4 to be a result of the Bank of Japan’s monetary policy. \par
\begin{figure}[htbp]
 \begin{minipage}[b]{0.45\linewidth}
  \centering
  \includegraphics[keepaspectratio, scale=0.6]{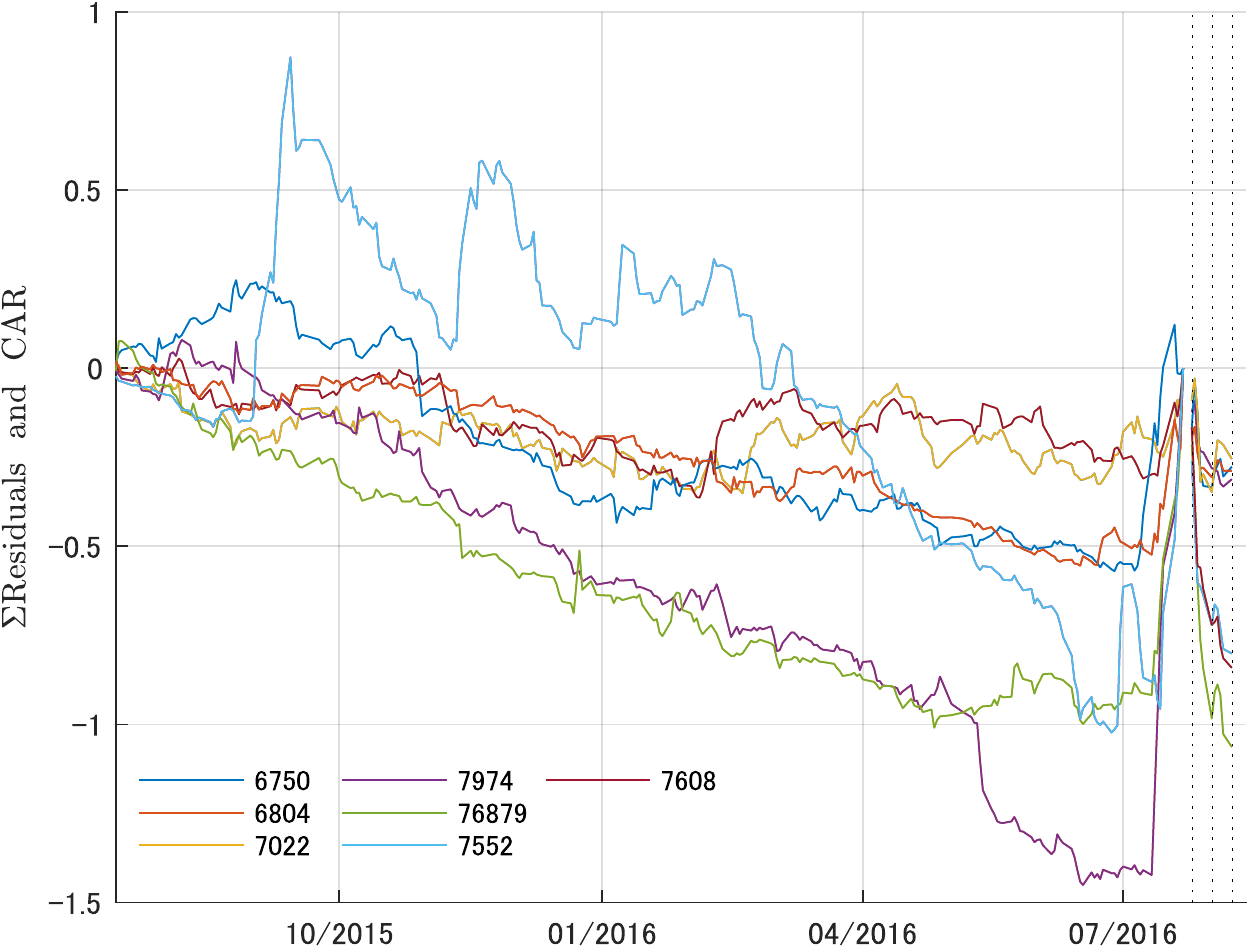}
  \subcaption{}
  \label{subFig:ResidGraph_A1(a)}
 \end{minipage}
 \begin{minipage}[b]{0.45\linewidth}
  \centering
  \includegraphics[keepaspectratio, scale=0.6]{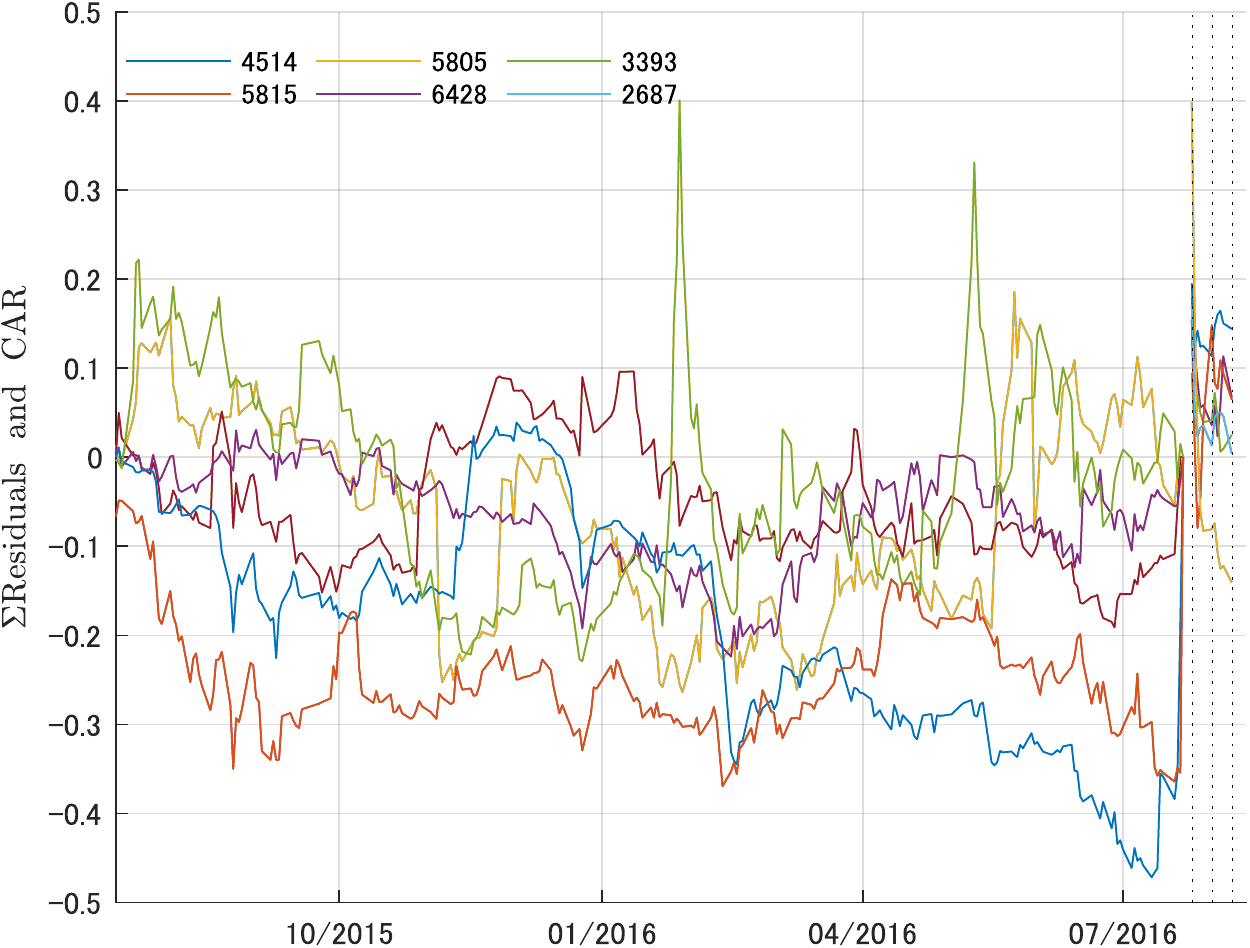}
  \subcaption{}
  \label{subFig:ResidGraph_A2(b)}
 \end{minipage}
 \caption{Cumulative residuals of estimated market model and cumulative abnormal returns plots for securities in (a) region A1 and (b) region A2 on Fig. \ref{Fig:SOM}.}
 \label{Fig:CumResidandCAR_A1A2} 
\end{figure}

Two abnormal returns in regions A1 and A2 cannot be explained by an earnings announcement. Regions A3 and A4 also include cells that cannot be explained by the same reason.
According to the submaps "DW Ratio", “$\alpha$" and "$\beta$, and three regions A1, A2, and A3 have points in common. That is, their Durbin-Watson ratios are less than 2, their $\alpha$s are relatively large, and their $\beta$s are almost 1.  
The Durbin-Watson ratio always lies between 0 and 4. If this ratio is substantially less than 2, it is evidence of positive residual serial correlation.
Fig.\ref{Fig:CumResidandCAR_A1A2} shows time series plots for the cumulative residuals of the estimated market model and the cumulative abnormal returns in regions A1 to A4.
Region A1 has seven securities (Tokyo Stock Exchange Codes 6750, 6804, 6879, 7022, 7552, 7608, and 7974). According to Fig.\ref{subFig:ResidGraph_A1(a)}, the stock returns of those seven securities rose before the event window.
Therefore, the cumulative residual of the seven securities exhibit distinctive shapes shown in Fig.\ref{subFig:ResidGraph_A1(a)}. Their Durbin-Watson Ratios are under 2; that is, they showed positive auto correlation.
It is well known that outliers can severely affect the parameter estimation of linear regression models because the OLS estimation scheme requires the summation of residuals to always equal zero.
Due to the positive spike return during the estimation window, the intercept in the market model of these six securities could have been overestimated.　　
This is because residuals are obtained by the observed value minus the predicted value, and the average of the residuals excluding those plus spikes must be minus.
The cause of the large plus returns was Pok\'{e}mon Go, a popular game for mobile devices.
Pok\'{e}mon Go was developed by Niantic and was initially released in the United States and other countries in July 2016. This game was the most downloaded application on the App Store during the first week after launch and was awarded five Guinness World Records by August 2016. 
Therefore, the equity price of Nintendo (7974) and its affiliated companies skyrocketed. 
The cumulative abnormal returns in region A1 are considered falsely detected because of outliers in the estimation window. \par
According to subgraph \ref{subFig:ResidGraph_A2(b)}, six securities in region A2 show a similar trend as region A1 shown in subgraph \ref{subFig:ResidGraph_A1(a)}.
Considering these securities, the election of the Tokyo Governor on July 31 may have caused a spike in returns. Ms. Koike, finally elected Tokyo’s Governor, offered some pledges. One pledge was to improve Tokyo’s landscape and another was to attract casinos to stimulate economic activity.
Out of six securities in A2, five securities (5805, 3393, 5815, 6428, and 2687) were companies expecting to see high growth because of the governor’s election promises. Ms. Koike's popularity was high, and the associated stock prices soared before the vote.
Therefore, it is probable that this detection was false and occurred for the same reasons as the case in region A1. \par
\begin{figure}[htbp]
  \begin{minipage}[t]{0.45\hsize}
    \centering
    \includegraphics[keepaspectratio, scale=0.6]{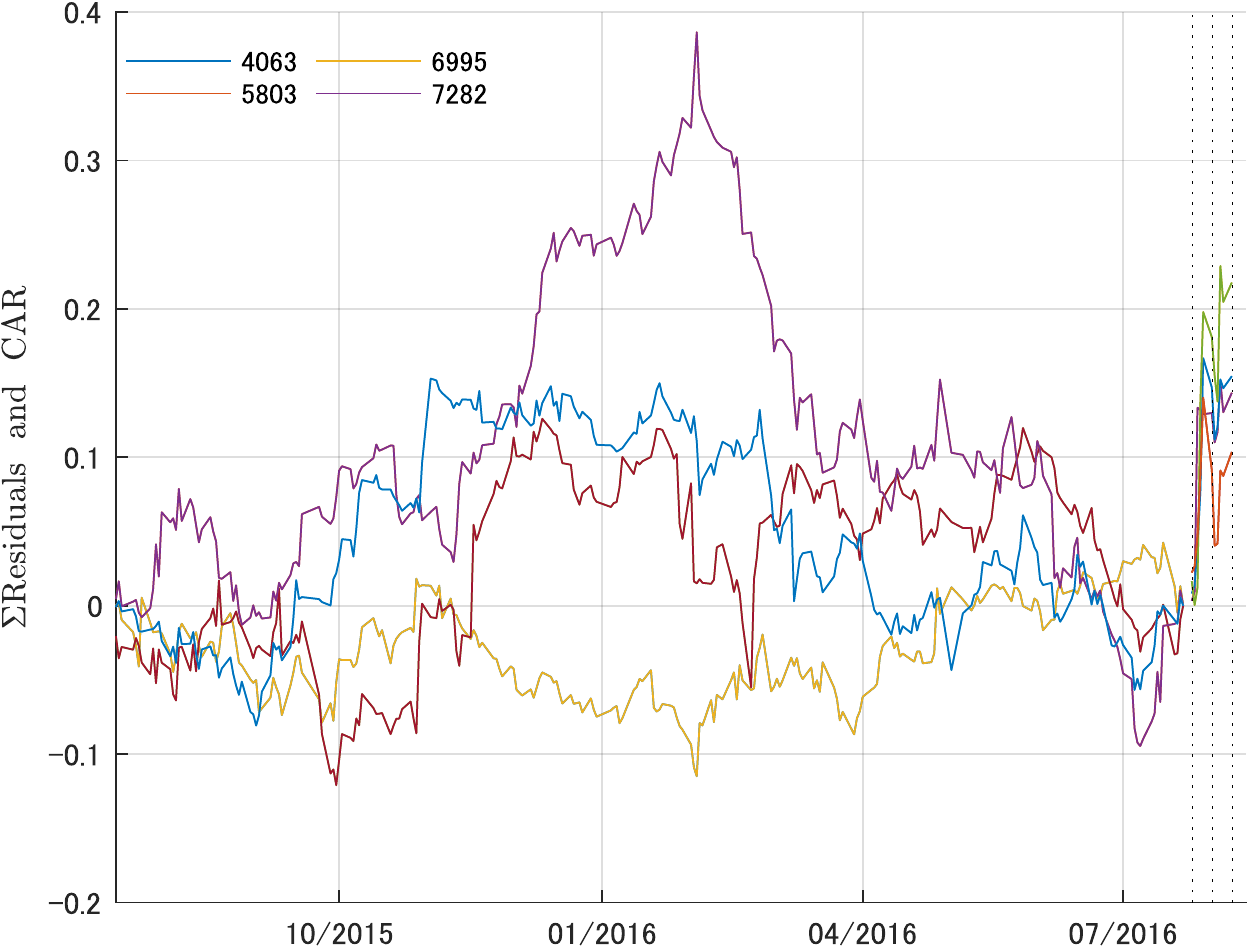}
    \subcaption{The left lattice in region A3}
    \label{subFig:ResidGraph_A3(a)}
  \end{minipage}
  \begin{minipage}[t]{0.45\linewidth}
    \centering
    \includegraphics[keepaspectratio, scale=0.6]{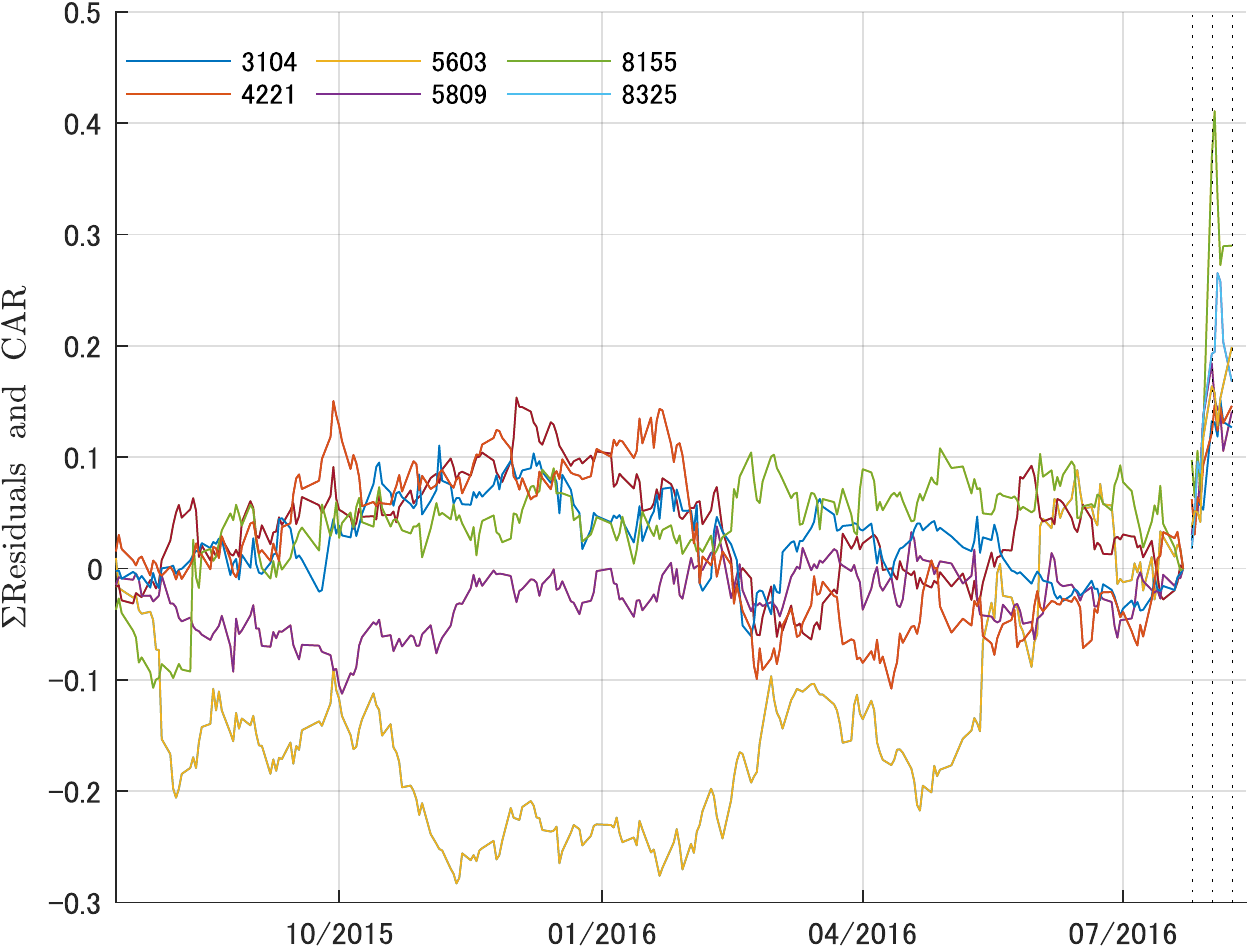}
    \subcaption{The middle lattice in region A3}
    \label{subFig:ResidGraph_A3(b)}
  \end{minipage}
  \begin{minipage}[t]{0.45\linewidth}
    \centering
    \includegraphics[keepaspectratio, scale=0.6]{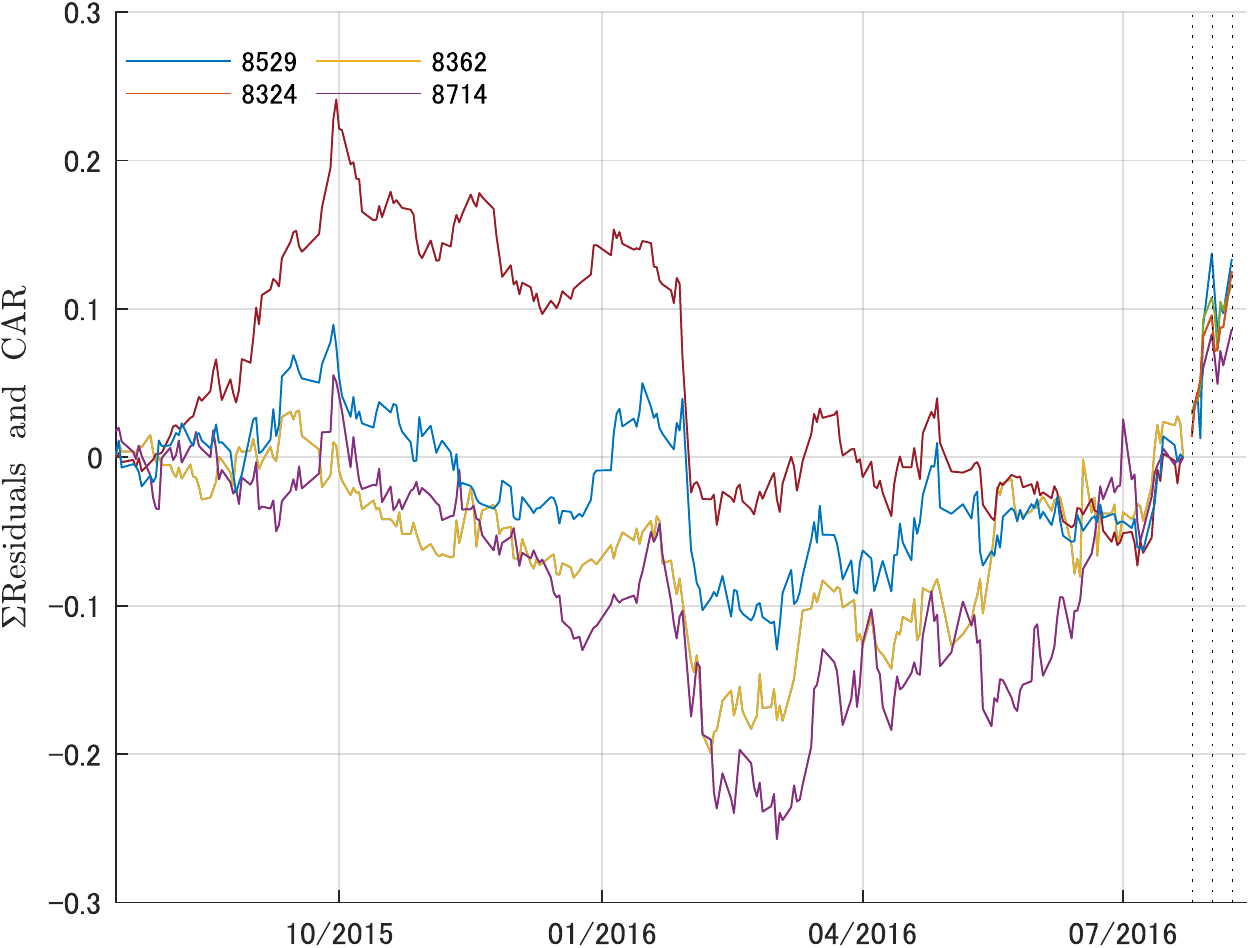}
    \subcaption{The right lattice in region A3 (The bottom lattice in region A4)}
    \label{subFig:ResidGraph_A3A4(c)}
  \end{minipage}
  \hfill\vspace{5pt}   %%%%%%%%%%%%%%%%%%%%%%%%%%%%%%%%%%%%%%%%%%%%%%%%%%%%%%
  \begin{minipage}[t]{.45\linewidth}  %{0.45\linewidth}      
   \centering
    \includegraphics[keepaspectratio, scale=0.6]{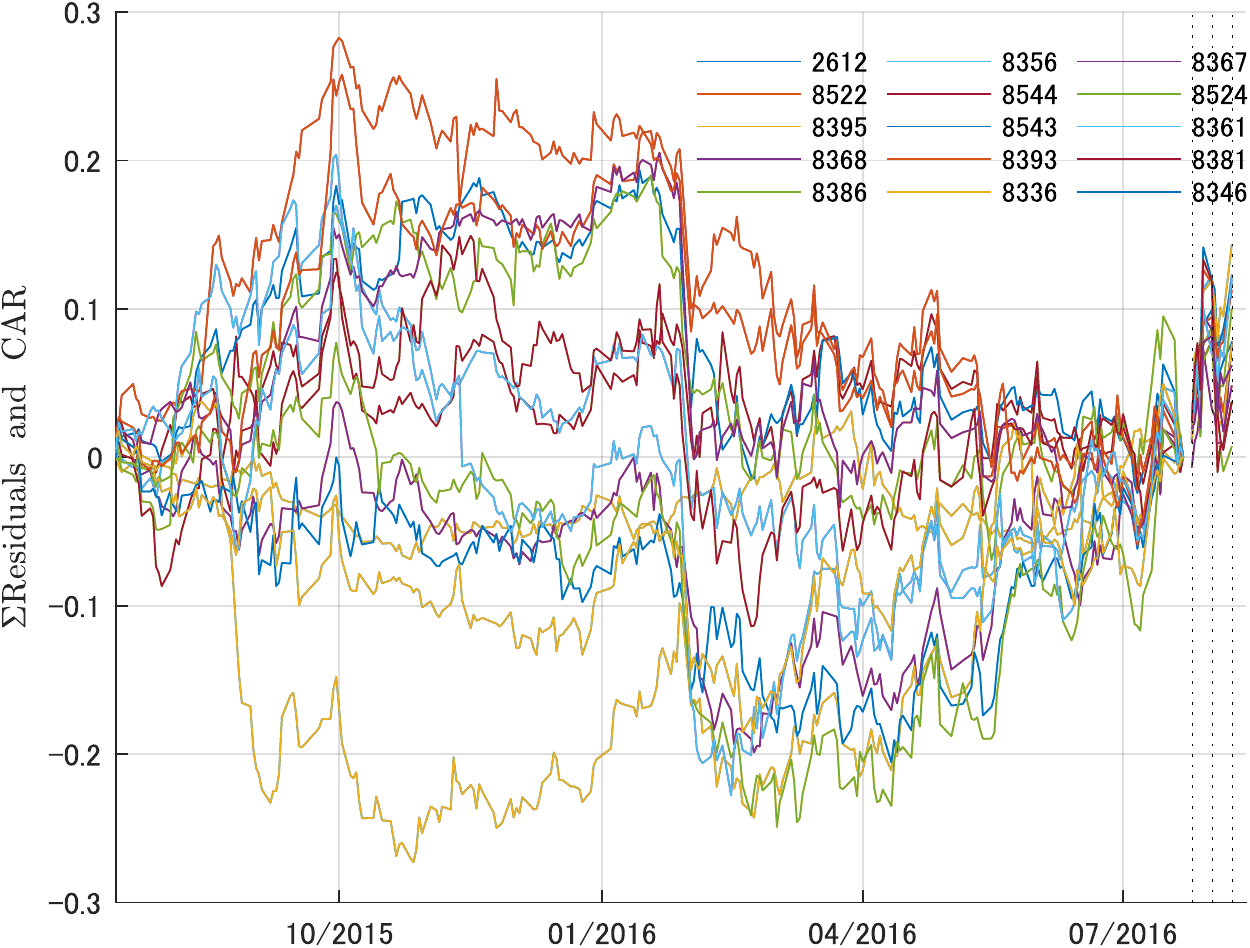}
    \subcaption{Region A4}
    \label{subFig:ResidGraph_A4(d)}
  \end{minipage}
  \caption{Cumulative residuals of estimated market model and cumulative abnormal returns plots for securities in regions A3 and A4 in Fig. \ref{Fig:SOM}.
  }\label{Fig:CumResidandCAR_A3A4}
\end{figure}

Subgraphs \ref{subFig:ResidGraph_A3(a)}, \ref{subFig:ResidGraph_A3(b)}, and \ref{subFig:ResidGraph_A3A4(c)} in Fig.\ref{Fig:CumResidandCAR_A3A4} show the cumulative residuals and cumulative abnormal return plots for each security belonging to three cells in region A3 obtained by the market model. Note that (c) lattice overlaps with the lattice in region A4. Subgraph \ref{subFig:ResidGraph_A4(d)} indicates the cumulative residuals and cumulative abnormal return plots of 15 securities in three cells that have low Durbin-Watson ratios belonging in region A4. 
The 14 securities of region A4 are composed of electric wire companies (5603, 5803, and 5809), banks (8325, 8529, 8362, 8324, and 8714), and companies whose parent company announced earnings in the event window.
Note that the shape of the cumulative residuals in Fig.\ref{subFig:ResidGraph_A1(a)} and Fig.\ref{subFig:ResidGraph_A4(d)} show a similar tendency despite our use of a dataset without residuals.   
This suggests that the detection of cumulative abnormal returns is susceptible to a spike in securities’ returns in the estimation window that is independent of the event window. Spike returns in the estimation window can have a negative effect on the preconditions of a linear regression model, which are the normality and homoscedasticity of a residual. The statistical test using an incorrect market model is unreliable.   
A linear regression model is often estimated by eliminating outliers to improve the model reliability. However, arbitrary exclusion of outliers in the estimation window will cause the statistical rationale of the event study to be lost.
With the conventional method, this problem cannot be avoided because it is impossible to link events and abnormal returns.
Our proposed method isolates the events causing abnormal returns, and it is possible to avoid this problem.\par
SOM can compress the multidimensional data while maintaining the topological relationship.
Securities placed in different cells on the two-dimensional map indicate that the market pricing system has nonlinearity.
From this perspective, within the same lattice, the behavior of securities is interpreted by the linear market model.
A specific linear market model with the same factor must be applied only to securities belonging to the same lattice.
This perspective is useful for the efficient modeling of accurate linear systems for securities prices.
For example, we cannot identify factor returns in arbitrage pricing theory (APT) \cite{roll1980empirical}, but we believe it is possible to identify them using our scheme.\par
The topological map analysis is a powerful tool that helps us to understand the nonlinear structure of a market as a system. Traditional pricing theories emphasize the structural understanding of the market and prefer simple line models. Therefore, we recognize the deviation between the actual returns and predictions of these models as abnormal returns. Traditional event studies are built on this concept. However, market interpretations that are too conceptual may interfere with an accurate understanding of the market system. Topological maps represent machine learning used for data mining analysis. We show the possibility of reducing risk with a combination of a machine learning scheme and traditional financial analysis.
Our scheme can be applied to traditional finance contexts. 
\section{Conclusion}
This work proposes a method for improving traditional event studies and empirical analysis. This scheme is a combination of a traditional event study using a linear regression market model and SOM, which is a type of topological map analysis in machine learning. SOM compresses multidimensional data to lower the dimension while holding its topological relation. Our proposed scheme can identify events correlated with abnormal returns.　Our scheme revealed the weaknesses in traditional event studies based on a linear regression market model. That is, there is a possibility of false detection of an abnormal return caused by a return spike during the estimation window. The interpretation of a flexible market structure using machine learning is expected to contribute to further development in the finance field.

% \medskip
%\section{Reference} 
\bibliographystyle{authordate1} % unsrt % plain %authordate1
%\bibliography{MyBibTakashi}
%\bibliography{Sample}
\bibliography{main.bbl}

\appendix
\section{Batch-learning SOM algorizm} \label{apdx:A}
%Initial weight vector $w_{i,j}$ is given by the equation 
The SOM has a two-layer structure; one is an input layer and the other is a competitive layer or output data. The input layer is composed of input vectors. 
The input vector corresponds to multidimensional data representing features of the analysis object (e.g., an active weight and t-value of a certain security in the event period.).
The competitive layer is composed of nodes in a dimension space lower than the input vector. Normally, the competitive layer is adopted as a two-dimensional plane. Each node has one vector of the same dimension as the input vector called a reference vector. This vector is updated according to the batch learning SOM algorithm described below, and each node has a reference vector similar to the neighboring node (This process is called learning.). Finally, each security is given a node label with a reference vector closest to its own feature vector. 
As a result, securities with similar characteristics are arranged on a two-dimensional map.
\begin{align}
\shortintertext{Suppose that the number of two dimensional lattice points, which are competing layers, are $I \times J$, initial weight vector $w_{I,j}$ is given by the equation}
w_{i,j} &= x_{\mathrm{ave}}+5\sigma_{1}b_{1}(\frac{i-I/2}{I})+5\sigma_{2}b_{2}(\frac{j-I/2}{J}). \\
\shortintertext{Here $x_{\mathrm{ave}}$ is the average of the input vector $x_{k}(k=1,2,\dots, N)$, $b_{1}$ and $b_{2}$ are first and second component vectors, and $\sigma_{1}$ and $\sigma_{2}$ are standard deviation, respectively. At the first step of learning, $x_{k}$ vectors are labeled on $W_{i',j'}$ with minimum Euclidean distance. At the next step, $W_{i,j}^{\mathrm{new}}$ vectors are updated following equation,  }
W_{i,j}^{\mathrm{new}} &= W_{i,j}^{\mathrm{old}} + \lambda(t)(\frac{\Sigma_{x_{k} \in S_{i,j}} x_{k}} {N_{i,j}} - W_{i,j}^{\mathrm{old}}).
\shortintertext{Here, $\lambda(t)$ is the learning coefficient with $0<\lambda(t)<1$, and $\xi(t)$ indicates the dispersion of the neighborhood of $W_{i,j}$. Neighboring ensemble $S_{i,j}$ satisfies two conditions as $i-\xi(t) \leq i' \leq i+\xi(t)$ and $ j-\xi(t) \leq j' \leq j+ \xi(t)$. $\lambda(t)$ and $\xi(t)$ are calculated following equation:}
\lambda(t) &= \max{ \{ 0.01, \lambda_{\mathrm{init}}(1-\frac{t}{T}) \} } \\
\beta(t) &= \max{ \{1, \xi_{\mathrm{init}-t} \} } 
\shortintertext{$\lambda_{\mathrm{init}}$ and $\xi_{\mathrm{init}}$ are initial parameters for the learning. $N_{i,j}$ is the element number of $S_{i,j}$, and $T$ is the iteration number. The learning result of each iteration is evaluated and $e(t)$ can be calculated as: }
e(t) &= \sum_{N}^{k=1}{ \{x_{k}-w_{i',j'} \}. }
\end{align}
\begin{table}[htbp]
\centering
\scalebox{0.8}[0.8]{ 
\begin{tabular}{lccl}
\multicolumn{1}{c}{Date} & \begin{tabular}[c]{@{}c@{}}Number of Companies\\ Earnings Announcement\end{tabular} & \multicolumn{1}{l}{\begin{tabular}[c]{@{}l@{}}Business Day Interval\\ from Event Date\end{tabular}} & \multicolumn{1}{c}{Market Events}                                                                   \\
\hline  \hline
25 Jul (Mon)             & 19                                                                                  & -5                                                                                                  &                                                                                                     \\
26 Jul (Tue)             & 41                                                                                  & -4                                                                                                  & FRB Federal Open Market Committee (1)                                                               \\
27 Jul (Wed)             & 70                                                                                  & -3                                                                                                  & FRB Federal Open Market Committee (2)                                                               \\
28 Jul (Thu)             & 195                                                                                 & -2                                                                                                  & BOJ Monetary Policy Meeting (1)                                                                     \\
29 Jul (Fri)             & 309                                                                                 & -1                                                                                                  & BOJ Monetary Policy Meeting (2)                                                                     \\
30 Jul (Sat)             & 0                                                                                   &                                                                                                     &                                                                                                     \\
31 Jul (Sun)             & 3                                                                                   &                                                                                                      & The election of heads in Tokyo                                                                      \\ 
{\bf 1 Aug (Mon)}      & 71                                                           &  0                                                                                                   & {\bf GPIF holdings disclosure}                                                                            \\
2 Aug (Tue)              & 65                                                                                  & +1                                                                                                  & \begin{tabular}[c]{@{}l@{}}An announcement on economic measures \\ (28.1 trillion yen)\end{tabular} \\
3 Aug (Wed)              & 89                                                                                  & +2                                                                                                  &                                                                                                     \\
4 Aug (Thu)              & 118                                                                                 & +3                                                                                                  &                                                                                                     \\
5 Aug (Fri)              & 218                                                                                 & +4                                                                                                  &                                                                                                     \\
6 Aug (Sat)              & 0                                                                                   &                                                                                                     &                                                                                                     \\
7 Aug (Sun)              & 0                                                                                   &                                                                                                     &                                                                                                     \\
8 Aug (Mon)              & 11                                                                                  & +5                                                                                                  &                                                                                                    
\end{tabular}}
%\end{minipage}
\caption{Market events around the event day}
\label{Tbl:Events}
\end{table}
\begin{table}[htbp]
\centering
\scalebox{0.8}[0.8]{ 
\begin{tabular}{crrrrrrrrrrr}
\multicolumn{1}{l}{\begin{tabular}[c]{@{}l@{}}Business Day Interval\\ from Event Date\end{tabular}} & \multicolumn{1}{c}{-5d} & \multicolumn{1}{c}{-4d} & \multicolumn{1}{c}{-3d} & \multicolumn{1}{c}{-2d} & \multicolumn{1}{c}{-1d} & \multicolumn{1}{c}{0d} & \multicolumn{1}{c}{+1d} & \multicolumn{1}{c}{+2d} & \multicolumn{1}{c}{+3d} & \multicolumn{1}{c}{+4d} & \multicolumn{1}{c}{+5d} \\ \hline
CAR                                                                                                 & 2.7\%                   & 4.5\%                   & 3.9\%                   & 4.2\%                   & 5.3\%                   & 6.5\%                 & 8.0\%                  & 6.7\%                  & 7.0\%                  & 8.1\%                  & 7.7\%                  \\
AR                                                                                                  & 2.7\%                   & 7.3\%                   & 4.4\%                   & 4.8\%                   & 5.5\%                   & 12.5\%                & 13.5\%                 & 6.2\%                  & 6.4\%                  & 16.0\%                 & 7.3\%                 
\end{tabular}}
\caption{Percentage of securities of CAR and AR detected in event window}
\label{Tbl:Percent}
\end{table}

\end{document}